\documentclass[%
reprint,
superscriptaddress,
showpacs,
amsmath,
amssymb,
aps,
pra,%
floatfix,
longbibliography,
]{revtex4-1}

\usepackage{graphicx}
\usepackage{dcolumn}
\usepackage{bm}
\usepackage{hyperref}
\usepackage{multirow}
\usepackage{tabularx}
\usepackage{amsmath,accents} 
\usepackage{textcase}

\RequirePackage{xspace}

\DeclareRobustCommand*{\etal}{et al.\xspace}

\newcommand{\SFO}{SrFe$_{12}$O$_{19}$}

\newcommand{\SFZSOx}{SrFe$_{12-2x}$(ZnSn)$_{x}$O$_{19}$}
\newcommand{\SFZnO}{SrFe$_{11.5}$Zn$_{0.5}$O$_{19}$}
\newcommand{\SFSnO}{SrFe$_{11.5}$Sn$_{0.5}$O$_{19}$}
\newcommand{\SFZnSnO}{SrFe$_{11}$(ZnSn)$_{0.5}$O$_{19}$}
\newcommand{\exptval}[1]{\left<{#1}\right>}

\newcolumntype{.}{D{.}{.}{-1}}

\begin{document}

\title{Correlation between site preference and magnetic properties
  of Zn-Sn-substituted strontium hexaferrite}

\author{Vivek~Dixit}
\affiliation{%
  Department of Physics and Astronomy, 
  Mississippi State University,
  Mississippi State, MS 39762, USA
}
\affiliation{%
  Center for Computational Sciences, 
  Mississippi State University, 
  Mississippi State, MS 39762, USA
}

\author{Dinesh~Thapa}
\affiliation{%
  Department of Physics and Astronomy, 
  Mississippi State University,
  Mississippi State, MS 39762, USA
}
\affiliation{%
  Center for Computational Sciences, 
  Mississippi State University, 
  Mississippi State, MS 39762, USA
}

\author{Bipin~Lamichhane}
\affiliation{%
  Department of Physics and Astronomy, 
  Mississippi State University,
  Mississippi State, MS 39762, USA
}
\affiliation{%
  Center for Computational Sciences, 
  Mississippi State University, 
  Mississippi State, MS 39762, USA
}

\author{Chandani~N.~Nandadasa}
\affiliation{%
  Department of Physics and Astronomy, 
  Mississippi State University,
  Mississippi State, MS 39762, USA
}
\affiliation{%
  Center for Computational Sciences, 
  Mississippi State University, 
  Mississippi State, MS 39762, USA
}

\author{Yang-Ki~Hong}
\affiliation{%
  Department of Electrical and Computer Engineering and MINT Center, 
  The University of Alabama,
  Tuscaloosa, AL 35487, USA
}

\author{Seong-Gon~Kim}
\email{Corresponding author: sk162@msstate.edu}
\affiliation{%
  Department of Physics and Astronomy, 
  Mississippi State University,
  Mississippi State, MS 39762, USA
}
\affiliation{%
  Center for Computational Sciences, 
  Mississippi State University, 
  Mississippi State, MS 39762, USA
}

\date{\today}

\begin{abstract}
  The site preference and magnetic properties of Zn, Sn and Zn-Sn
  substituted M-type strontium hexaferrite (SrFe$_{12}$O$_{19}$) have
  been investigated using first-principles total energy calculations
  based on density functional theory.  The site occupancy of
  substituted atoms were estimated by calculating the substitution
  energies of different configurations. The distribution of different
  configurations during the annealing process at high temperature was
  determined using the formation probabilities of configurations to
  calculate magnetic properties of substituted strontium
  hexaferrite. We found that the magnetization and magnetocrystalline
  anisotropy are closely related to the distributions of Zn-Sn ions
  on the five Fe sites.  Our calculation show that in
  SrFe$_{11.5}$Zn$_{0.5}$O$_{19}$, Zn atoms prefer to occupy $4f_1$,
  $12k$, and $2a$ sites with occupation probability of 78\%, 19\% and
  3\%, respectively, while in SrFe$_{11.5}$SnO$_{19}$, Sn atoms occupy
  the $12k$ and $4f_2$ sites with occupation probability of 54\% and
  46\%, respectively.  We also found that in
  SrFe$_{11}$Zn$_{0.5}$Sn$_{0.5}$O$_{19}$, (Zn,Sn) atom pairs prefer
  to occupy the ($4f_1$, $4f_2$), ($4f_1$, $12k$) and ($12k$, $12k$)
  sites with occupation probability of 82\%, 8\% and 6\%,
  respectively.  Our calculation shows that the increase of
  magnetization and the reduction of magnetic anisotropy in Zn-Sn
  substituted M-type strontium hexaferrite as observed experimentally
  is due to the occupation of (Zn,Sn) pairs at the ($4f_1$, $4f_2$)
  sites.
\end{abstract}

\maketitle

\section{Introduction}
\label{sec:intro} 
Hexagonal strontium hexaferrite \SFO\ (SFO), along with other $M$-type
hexaferrites $X$Fe$_{12}$O$_{19}$ ($X$ = Sr, Ba, Pb), has large
saturation magnetization, high coercivity, and excellent chemical
stability \cite{Pang2010, Davoodi2011, Ashiq2012}.  Several
experimental studies of substituted hexaferrite have been performed
where magnetic properties are tailored to fit specific applications by
the partial substitution of Fe in its crystallographic sites by
divalent, trivalent, tetravalent, and divalent-tetravalent combination
of metal atoms. The substitution of rare-earth elements such as Pr
\cite{OUNNUNKAD2006, WANG2005}, La \cite{WANG2004, SEIFERT2009}, Sm
\cite{WANG2001}, and Nd \cite{WANG2002} in M-type hexaferrite has
shown to enhance the coercivity without much reduction in
magnetization.  Rare earth elements increase the spin-orbit coupling
which in turn strengthen the magnetocrystalline anisotropy, hence the
coercivity.  However, reducing the use of rare-earth elements is
highly desired for economical reasons. The substitution of Al leads to
enhancement in coercivity \cite{Dixit2015a, Albanese1974,
  Awawdeh2014}. Wang \etal\ \cite{WANG2012} have reported coercivity
values as high as 17.6~kOe for Al-substituted strontium hexaferrite.
Even higher values of coercivity (21.3~kOe) are possible by the double
substitution of Ca and Al atoms \cite{TRUSOV2018}, which is even
higher than the coercivity of Nb-based magnets.  The substitution of
Fe by the trivalent metals such as Al, Ga, and Cr leads to an increase
in coercivity and magnetocrystalline anisotropy (MAE) with reduction
in magnetization \cite{Awawdeh2014}.  In particular, the substitution
of divalent-tetravalent pairs such as Zn-Nb \cite{FANG2004}, Zn-Sn
\cite{FANG1998129, MENDOZASUAREZ:2001, Ghasemi:2010, Ghasemi2011}, and
Sn-Mg \cite{Davoodi2011} result in significant enhancement in
saturation magnetization with rapid reduction in coercivity.

Theoretical studies on pure and substituted M-type hexaferrite
have also been performed.  Fang \etal\ investigated the electronic
structure of strontium hexaferrite using density-functional theory
(DFT) \cite{Fang2003}.  Park \etal\ have calculated the exchange
interaction of strontium hexaferrite from the differences of the total
energy of different collinear spin configurations
 \cite{Park2014}. Magnetism in La-substituted strontium hexaferrite has
been studied using DFT \cite{Kupferling2005, Novak2005a}. Zn-Sn
substituted strontium hexaferrite has been studied by Liyanage \etal\
 \cite{Liyanage2013}.  The site occupancy and magnetic properties of
Al, In, and Ga-substituted strontium hexaferrite has been investigated
by Dixit \etal \cite{Dixit2015a, Dixit2015b}.

In this work, we used first-principles total-energy calculations to
investigate the relationship between the site occupation and magnetic
properties of substituted strontium hexaferrite,
SrFe$_{12-x}M_{x}$O$_{19}$ with $M = \mathrm{Zn}$ or Sn and
SrFe$_{12-2x}$(ZnSn)$_x$O$_{19}$ with $x = 0.5$.  The Boltzmann
distribution function was used to determine the formation
probabilities of various configurations at a typical annealing
temperature (1000~K) of strontium hexaferrite, which was further used
to compute the weighted average of various magnetic properties.  We
show that our model predicts an increase of saturation magnetization
as well as a decrease in magnetic anisotropy energy (MAE) of \SFZnSnO\
compared to the pure $M$-type SFO in good agreement with the
experimental observations \cite{Ghasemi2010, Ghasemi2011}.

\begin{figure}[tbp]
  \centering
  \includegraphics[scale=0.5,keepaspectratio=true]{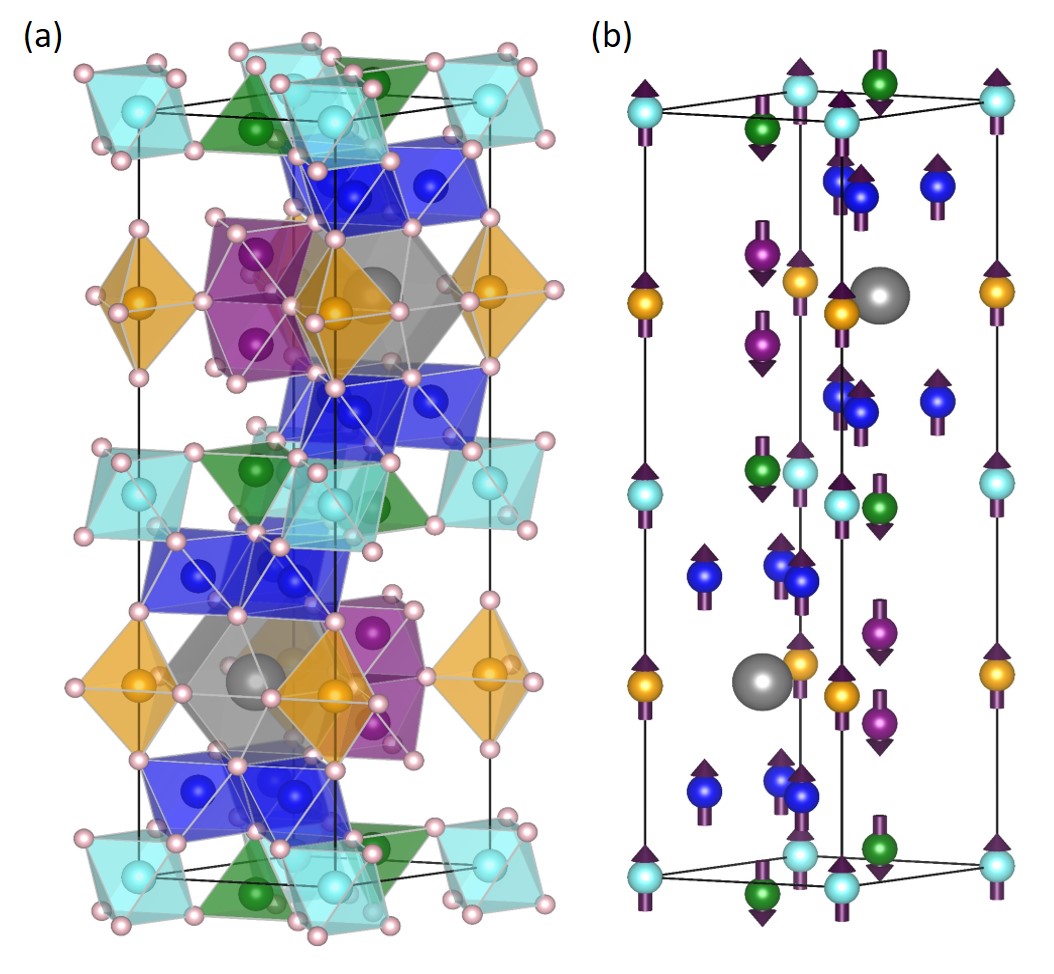}
  \caption{\label{fig:sfo}
    (a) One double formula unit cell of \SFO. Two large gray spheres
    are Sr atoms and small pink spheres are O atoms.
    Colored spheres enclosed by polyhedra formed
    by O atoms represent Fe$^{3+}$ ions in different inequivalent
    sites: $2a$ (cyan), $2b$ (orange), $4f_1$ (green),
    $4f_2$ (purple), and $12k$ (blue).
    (b) A schematic diagram of the lowest-energy
    spin configuration of Fe$^{3+}$ ions of \SFO.  The arrows
    represent the local magnetic moment at each atomic site.  }
\end{figure} 

\section{Methods}
\label{sec:method}

SFO belongs to space group $P6_{3}/mmc$ (No.~194) and has a hexagonal
magnetoplumbite crystal structure \cite{OBRADORS1988} whose double
formula unit cell containing 64 atoms is shown in Fig.~\ref{fig:sfo}.
The unit cell structure of SFO consists of ten oxygen layers and the
Fe$^{+3}$ ions occupy five crystallographically inequivalent sites:
three octahedral sites ($2a$, $12k$, and $4f_2$), one tetrahedral site
($4f_1$), and one trigonal bipyramid site ($2b$) as indicated by the
coordination polyhedra in Fig.~\ref{fig:sfo}(a). As a ferrimagnetic
material, SFO has 16 Fe$^{+3}$ ions with spins in the majority
direction ($2a$, $2b$, and $12k$ sites) and 8 Fe$^{+3}$ ions with
spins in the minority direction ($4f_1$ and $4f_2$ sites) as shown in
Fig.~\ref{fig:sfo}(b).  First-principles total-energy calculations
were performed to determine the site preference of Zn and Sn atoms in
$M$-type Sr-hexaferrite. Total energies and forces were calculated
using density-functional theory (DFT) with projector augmented wave
(PAW) potentials as implemented in VASP \cite{Kresse1996,
  Kresse1999}. The exchange correlation effect was described using the
Perdew-Burke-Ernzerhof (PBE) within generalized gradient approximation
(GGA) \cite{Perdew1996}.  All calculations were spin polarized
according to the ground state ferrimagnetic ordering of Fe spins
\cite{Fang2003, Gorter1957}.  Electronic wave functions were expanded
in a plane-wave basis with an energy cutoff of 520~eV.  Reciprocal
space was sampled with a $7\times 7\times 1$ Monkhorst-Pack mesh
\cite{Monkhorst1976} with a Fermi-level smearing of 0.2~eV applied
through the Methfessel-Paxton method \cite{Methfessel1989} for
relaxations and the tetrahedron method \cite{Bloch1994tetra} for
static calculations.  Full geometrical optimization was performed to
relax the positions of ions, cell shape, and cell volume until the
largest force component on any ion was less than 0.01~eV/\AA.  Since
the on-site Coulomb interactions are particularly strong for localized
$d$ electrons, we employed GGA+U method in the simplified rotationally
invariant approach in order to avoid the self-interaction error in
localized electron state of Fe-$3d$ \cite{Dudarev1998}.  Based
on our previous study \cite{Liyanage2013}, $U_{\text{eff}}$ for Fe
atoms was set to 3.7~eV.  The $U_{\text{eff}}$ for Zn
($3d^{10}4s^{2}$), Sn ($5s^{2}5p^{2}$), Sr ($5s^{2}$), and O
($2s^{2}2p^{4}$) were set to zero. Drawings in Fig.~\ref{fig:sfo} are
produced using VESTA code \cite{Momma:db5098}.

The magnetic properties of SFO can be modified by substitution of
foreign atoms for Fe. There are five crystallographic inequivalent Fe
sites in SFO. When foreign atoms are substituted in a SFO unit cell,
there can be several energetically different configurations. Due to
the ferrimagnetic nature of SFO, the magnetization of substituted SFO
strongly depends on site preference of the substituted atoms. In order
to investigate the effect of substitution on the magnetic properties,
it is imperative to understand site preference of substituted
atoms. The site preference of the substituted atom can be determined
by calculating the substitution energy. The substitution energy
$E_{\text{sub}}[i]$ for configuration $i$ at 0~K is given by
\begin{equation} \label{eq:Esub}
  E_{\text{sub}}[i] = E_{\text{SFXO}}[i] - E_{\text{SFO}} 
  -\sum_{\alpha}n_{\alpha}\epsilon_{\alpha} 
\end{equation}
where $E_{\text{SFXO}}[i]$ is the total energy per unit cell of
substituted SFO in configuration $i$, while $E_{\text{SFO}}$ is the
total energy per unit cell of pure SFO and $\epsilon_{\alpha}$ is the
total energy per atom for element $\alpha$ ($\alpha$ = Zn, Sn and Fe)
in its most stable crystal structure. Zn belongs to space group
$P6_{3}/mmc$ (No.~194) has a hexagonal crystal structure while Sn
belongs to $Fd\bar{3}m$ (No.~227) with cubic crystal system.
$\epsilon_{Zn}$, $\epsilon_{Sn}$ and $\epsilon_{Fe}$ were found to be
$-0.789$~eV, $-3.835$~eV and $-8.461$~eV, respectively.  $n_\alpha$ is
the number of atoms of type $\alpha$ added or removed; if one atom is
added then $n_{\alpha} = +1$ while $n_\alpha = -1$ when one atom is
removed.

The magnetocrystalline anisotropy energy, $E_a$, was also calculated.
$E_a$ in the present case, is defined as the difference between the
two total energies where the spin quantization axes are aligned along
two different directions \cite{Ravindra:1999}:
\begin{equation}
  \label{eq:E_a}
  E_{a} = E_{(100)} - E_{(001)}
\end{equation}
where, $E_{(100)}$ is the total energy with spin quantization axis in
the magnetically hard plane and $E_{(001)}$ is the total energy with
spin quantization axis in the magnetically easy axis.  The total
energies in Eq.~(\ref{eq:E_a}) are computed by the non-self-consistent
calculations, where the spin densities are kept
constant \cite{Daalderop:1990}. 

The uniaxial magnetic anisotropy constant, $K_1$, can be computed
as \cite{Munoz:2013, Smit:1959}:
\begin{equation}
  K_{1} = \frac{E_a}{V\sin^{2}\theta}
\end{equation} 
where $V$ is the equilibrium volume of the unit cell, and $\theta$ is
the angle between the two spin quantization axis orientations
(90$^\circ$ in the present case).  The anisotropy field, $H_a$, which
is related to the coercivity can be expressed as \cite{Kittel:1949}:
\begin{equation}
  H_a = \frac{2K_1}{M_s}
\end{equation} 
where $K_1$ is the magnetocrystalline anisotropy constant and $M_s$ is
the saturation magnetization.

When the separation between $E_{sub}$ of different configurations is
not too big compared to the thermal energy during the synthesis of
these hexaferrites at a high annealing temperature ($\gtrsim$ 1000~K),
we can expect the site preference of substituted atoms to change at
such an elevated temperature. This change in the site occupation
preference can be modeled using Maxwell-Boltzmann distribution.  The
site occupation probability or the formation probability $P_{i}(T)$ of
configuration $i$ at temperature $T$ is given by
\begin{equation}
  \label{eq:P(i)}
  P_{i}(T) = \frac{g_{i}\exp(-\Delta G_{i}/k_{B}T)}
  {\sum_{j} g_{j}\exp(-\Delta G_{j}/k_{B}T)}
\end{equation} 
\begin{equation} \label{eq:G(i)}
  \Delta G_{i} = \Delta E_{\text{i}} + P\Delta V_{i} - T\Delta S_{i} 
\end{equation}
\begin{equation} \label{eq:S(i)}
  \Delta S_{i} = k_{B}\ln(g_{i}) - k_{B}\ln(g_{0}) 
\end{equation}
where $\Delta G_{i}$, $\Delta E_{\text{i}}$, $\Delta V_{i}$, and
$\Delta S_i$ are the change in free energy, substitution energy, unit
cell volume and entropy of the configuration $i$ relative to the
ground state configuration.  $P$, $k_{\text{B}}$ and $g_{i}$ are the
pressure, Boltzmann constant and multiplicity of configuration $i$.
$g_{0}$ is the multiplicity of the ground state configuration.  In our
earlier work, we considered $\Delta S_i$ to be the same for all
configurations \cite{Dixit2017}.  Eq.~(\ref{eq:S(i)}) improves the model
by explicit calculation of change in entropy with respect to the most
stable configuration \cite{Gilmore2015}. 

Therefore, when the formation probability of higher energy
configurations become non-negligible at the annealing temperature, it
can be concluded that in a sample of substituted SFO there will not be
a single configuration but a distribution of several configurations.
Any physical quantity of a sample of SFO will then be a weighted
average of respective quantity in different configurations:
\begin{equation}
  \label{eq:prop}
  \exptval{Q} = \sum\limits_{i} P_{\text{1000K}}(i)\cdot Q_{i}
\end{equation} 
where $P_{\text{1000K}}(i)$ and $Q_{i}$ are the formation probability
at 1000~K and a physical quantity $Q$ of the configuration $i$.
During the annealing process when the material is maintained at high
temperature, substituted atoms have sufficient energy to overcome the
local energy barriers and acquire energetically the various
configurations that have sufficient formation probability. We note
that the substituted SFO considered in the present work loses its
magnetic properties at a typical annealing temperature (1000~K or
higher) that is near or above its Curie temperature.  When the
material is cooled down to a low temperature below the critical
temperature, it regains the magnetic properties while the
configurations are locked in those with higher substitution energy due
to energy barriers between them.  Consequently, the weighted average
calculated by Eq.~(\ref{eq:prop}) is the material's low temperature
property even though $P_{\text{1000K}}$ is used for computation.

\section{Results and Discussion}
\label{sec:Results}

In this work, two types of substitutions have been studied. In the
first case one Zn or Sn atom was substituted in a SFO unit cell. In
the second case, a pair of Zn and Sn atoms were substituted in a SFO
unit cell.  Although there are 24 Fe atoms in SFO unit cell, the
application of crystallographic symmetry operations shows that many of
these Fe sites are equivalent and leaves only five inequivalent
sites. Therefore, one foreign atom can be substituted in five
different ways, giving rise to five different configurations. We label
these inequivalent configurations using the crystallographic name of
the Fe site: $[2a]$, $[2b]$, $[4f_{1}]$, $[4f_{2}]$, and $[12k]$.
$E_{\text{sub}}$ corresponding to the substitution of one Zn at
different inequivalent Fe sites is given in
Table~\ref{tab:SFZnOx0.5}. The lowest $E_{\text{sub}}$ was found when
a Zn atom was substituted at $4f_{1}$ site, followed by the $[2a]$ and
$[12k]$ cases.  For the case of a single Sn atom substitution
(Table~\ref{tab:SFSnOx0.5}) the lowest $E_{\text{sub}}$ was found to
be for the substitution at the $4f_{2}$ site, and the second lowest
$E_{\text{sub}}$ was corresponding to the substitution at the $12k$
site.  A high multiplicity of the $12k$ site and low $E_{\text{sub}}$
indicate that at higher temperatures the $12k$ site is very likely to
be occupied.  In the second case, where a pair of Zn and Sn atoms
were substituted in the SFO unit cell, there can be $5\times5=25$
different configurations. $E_{\text{sub}}$ corresponding to all these
configurations are listed in Table~\ref{tab:SFZnSnOx0.5}. The lowest
$E_{\text{sub}}$ was found to be the configuration where Zn goes to
the $4f_1$ while Sn occupies the $4f_2$ site.

\begin{table*}[tbp]
  \caption{\label{tab:SFZnOx0.5} Physical properties of inequivalent
    configurations of SrFe$_{12-x}$Zn$_{x}$O$_{19}$ with $x=0.5$:
    multiplicity ($g$), substitution energy ($E_{\text{sub}}$), volume
    of the unit cell ($V$), total magnetic moment ($m_{\text{tot}}$),
    saturation magnetization ($\sigma_s$), magnetocrystalline
    anisotropy energy ($E_a$), uniaxial magnetic anisotropy constant
    ($K_1$), anisotropy field ($H_a$), and the formation probability
    at 1000~K ($P_{\text{1000K}}$).  All values are for a double
    formula unit cell containing 64 atoms. }
  \centering
  \begin{ruledtabular}
    \begin{tabular*}{\hsize}{cccccccccccr}
      config & $g$
      &  $E_{\text{sub}}$ (eV) & $V$ ($\AA^{3}$)
      & $m_{\text{tot}}$ ($\mu_{\text{B}}$) & $\sigma_{s}$ (emu/g)
      & $E_a$ (meV) & $K_1$ (kJ/m$^{3}$) & $H_a$ (kOe) & $P_{\text{1000K}}$ \\
      \hline
      $[4f_{1}]$ & 4  & -1.44 & 706.61
      & 44 & 115.01 & 0.83 & 188.20 & 6.37 & 0.794 \\
      $[2a]$     & 2  & -1.27 & 705.15
      & 34 & 88.74 & 0.90 & 204.49 & 8.95 & 0.025 \\ 
      $[12k]$    & 12 & -1.13 & 706.61
      & 34 & 88.70 & 0.80 & 181.39 & 7.96 & 0.180 \\
      $[2b]$     & 2  & -0.86 & 708.08
      & 36 & 94.13 & 0.63 & 142.55 & 5.91 & 0.000 \\ 
      $[4f_{2}]$ & 4  & -0.63 & 708.14
      & 44 & 115.15 & 0.85 & 192.31 & 6.52 & 0.000 
    \end{tabular*}
  \end{ruledtabular}
\end{table*}

\begin{table*}[tbp]
  \caption{\label{tab:SFSnOx0.5} Physical properties of inequivalent
    configurations of SrFe$_{12-x}$Sn$_{x}$O$_{19}$ with $x=0.5$:
    multiplicity ($g$), substitution energy ($E_{\text{sub}}$), volume
    of the unit cell ($V$), total magnetic moment ($m_{\text{tot}}$),
    saturation magnetization ($\sigma_s$), magnetocrystalline
    anisotropy energy ($E_a$), uniaxial magnetic anisotropy constant
    ($K_1$), anisotropy field ($H_a$), and the formation probability
    at 1000~K ($P_{\text{1000K}}$).  All values are for a double
    formula unit cell containing 64 atoms. }
  \centering
  \begin{ruledtabular}
    \begin{tabular*}{\hsize}{cccccccccccr}
      config & $g$
      &  $E_{\text{sub}}$ (eV) & $V$ ($\AA^{3}$)
      & $m_{\text{tot}}$ ($\mu_{\text{B}}$) & $\sigma_{s}$ (emu/g)
      & $E_a$ (meV) & $K_1$ (kJ/m$^{3}$) & $H_a$ (kOe) & $P_{\text{1000K}}$ \\
      \hline
      $[4f_{2}]$ &  4 & -2.42 & 716.94
      & 44 & 112.37 & 0.66 & 147.49 & 5.06 & 0.462 \\
      $[12k]$    & 12 & -2.25 & 716.65
      & 36 &  91.95 & 0.83 & 185.56 & 7.77 & 0.538 \\
      $[2b]$     &  2 & -1.78 & 712.67
      & 34 &  86.74 & 0.66 & 148.38 & 6.55 & 0.000 \\
      $[4f_{1}]$ &  4 & -1.51 & 715.41
      & 44 & 112.38 & 0.18 & 40.31 & 1.38 & 0.000 \\
      $[2a]$     &  2 & -1.51 & 716.45
      & 35 &  90.65 & 0.90 & 201.27 & 8.55 & 0.000 \\
    \end{tabular*}
  \end{ruledtabular}
\end{table*}

\begin{table*}[tbp]
  \caption{\label{tab:SFZnSnOx0.5} Physical properties 
    of inequivalent configurations of SrFe$_{12-2x}$(ZnSn)$_x$O$_{19}$
    with $x = 0.5$:  multiplicity ($g$), substitution energy ($E_{\text{sub}}$), volume
    of the unit cell ($V$), total magnetic moment ($m_{\text{tot}}$),
    saturation magnetization ($\sigma_s$), magnetocrystalline
    anisotropy energy ($E_a$), uniaxial magnetic anisotropy constant
    ($K_1$), anisotropy field ($H_a$), and the formation probability
    at 1000~K ($P_{\text{1000K}}$).  All values are for a double
    formula unit cell containing 64 atoms. }
  \centering
  \begin{ruledtabular}
    \begin{tabular*}{\hsize}{ccccccccccc}
      config & $g$
      &  $E_{\text{sub}}$ (eV) & $V$ ($\AA^{3}$)
      & $m_{\text{tot}}$ ($\mu_{\text{B}}$) & $\sigma_{s}$ (emu/g)
      & $E_a$ (meV) & $K_1$ (kJ/m$^{3}$) & $H_a$ (kOe) & $P_{\text{1000K}}$ \\
      \hline
      $[4f_{1},4f_{2}]$ & 16 & -5.326 & 717.32
      & 50 & 127.13 & 0.70 & 156.35 & 4.72 & 0.819 \\
      $[2b,4f_{2}]$    &  8 & -5.124 & 719.68
      & 40 & 101.97 & 0.58 & 129.12 & 4.88 & 0.020 \\
      $[4f_{1},2a]$    &  8 & -5.065 & 716.49
      & 40 & 101.68 & 0.82 & 183.36 & 6.91 & 0.010 \\
      $[2a,4f_{2}]$    &  8 & -4.991 & 714.67
      & 40 & 101.72 & 0.82 & 183.83 & 6.91 & 0.004 \\ 
      $[4f_{1},12k]$   & 48 & -4.940 & 717.54
      & 40 & 101.72 & 0.61 & 136.21 & 5.14 & 0.084 \\
      $[12k,12k]$       & 132 & -4.736 & 715.33
      & 30 &  76.17 & 0.77 & 172.46 & 8.67 & 0.060 \\
      $[2a,12k]$       & 24 & -4.697 & 715.08
      & 30 &  76.28 & 0.83 & 185.97 & 9.33 & 0.001 \\
      $[12k,4f_{2}]$   & 48 & -4.663 & 717.04
      & 40 & 101.58 & 0.63 & 140.77 & 5.32 & 0.003 \\ 
      $[4f_{1},2b]$    &  8 & -4.650 & 715.16
      & 40 & 101.72 & 0.49 & 109.78 & 4.13 & 0.000 \\
      $[12k,2a]$       & 24 & -4.366 & 718.08
      & 30 &  76.17 & 0.71 & 158.42 & 7.99 & 0.000 \\
      $[12k,2b]$       & 24 & -4.207 & 715.13
      & 30 &  76.15 & 0.49 & 109.78 & 5.52 & 0.000 \\
      $[2a,4f_{1}]$    &  8 & -4.206 & 714.79
      & 40 & 101.68 & 0.90 & 210.73 & 7.59 & 0.000 \\ 
      $[2a,2a]$         &  2 & -4.156 & 720.35
      & 30 &  76.26 & 0.85 & 189.05 & 9.56 & 0.000 \\ 
      $[2b,12k]$ & 24  & -4.119 & 721.64
      & 30 &  76.39 & 0.58 & 128.77 & 6.51 & 0.000 \\
      $[4f_{2},2b]$ & 8  & -4.082 & 716.07
      & 40 & 101.73 & 0.53 & 118.59 & 4.47 & 0.000 \\
      $[4f_{1},4f_{1}]$   &  12 & -3.946 & 720.60
      & 50 & 127.17 & 0.78 & 173.43 & 5.26 & 0.000 \\
      $[12k,4f_{1}]$     & 48 & -3.932 & 717.21
      & 40 & 101.59 & 0.84 & 187.65 & 7.09 & 0.000 \\
      $[4f_{2},4f_{2}]$   & 12 & -3.876 & 721.34
      & 50 & 127.18 & 0.72 & 159.92 & 4.85 & 0.000 \\
      $[2a,2b]$  & 4  & -3.726 & 717.56
      & 30 & 76.58 & 0.79 & 176.39 & 8.83 & 0.000 \\
      $[2b,2a]$  & 4  & -3.706 & 717.50
      & 30 & 76.58 & 0.79 & 175.79 & 8.85 & 0.000 \\
      $[4f_{2},12k]$ & 48 & -3.586 & 723.09
      & 40 & 101.86 & 0.75 & 166.12 & 6.31 & 0.000 \\
      $[4f_{2},2a]$ & 8 & -3.493 & 724.60
      & 40 & 101.73 & 1.12 & 246.98 & 9.41 & 0.000 \\
      $[2b,4f_{1}]$ & 8 & -3.297 & 722.28
      & 40 & 101.83 & 0.68 & 150.84 & 5.73 & 0.000 \\ 
      $[2b,2b]$ & 2 & -3.086 & 690.64
      & 30 & 76.36 & 0.41 & 95.11 & 4.60 & 0.000 \\ 
      $[4f_{2},4f_{1}]$ & 16 & -3.073 & 723.67
      & 50 & 127.21 & 0.78 & 172.69 & 5.26 & 0.000 
    \end{tabular*}
  \end{ruledtabular}
\end{table*}

\begin{figure}[tbp]
  \centering
  \begin{tabular}{c}
    \includegraphics[scale=0.35,keepaspectratio=true]{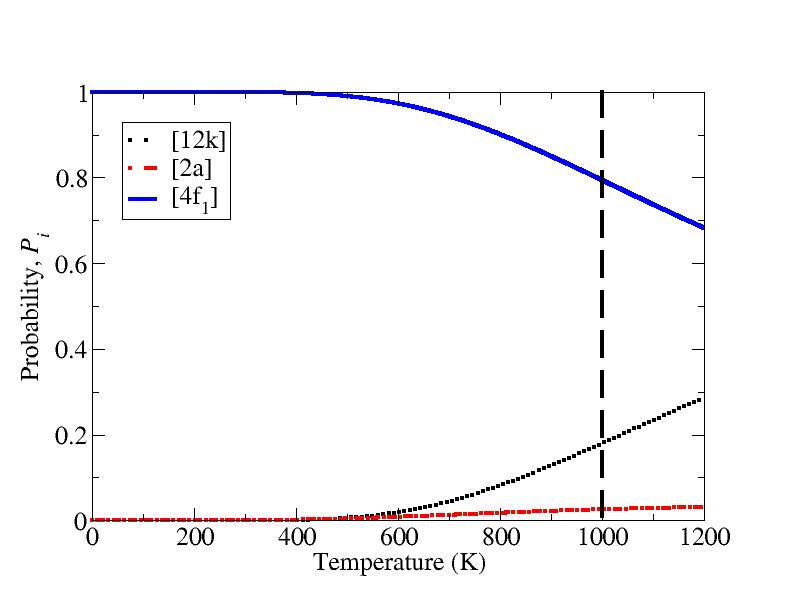}
  \end{tabular}
  \caption{\label{fig:Zn} Temperature dependence of the formation
    probability of different configurations of
    SrFe$_{11.5}$Zn$_{0.5}$O$_{19}$.  The configurations with
    negligible probability are not shown.  The vertical dotted line
    indicates the annealing temperature of 1000~K.  }
\end{figure}

\begin{figure}[tbp]
  \centering
  \begin{tabular}{c}
    \includegraphics[scale=0.35,keepaspectratio=true]{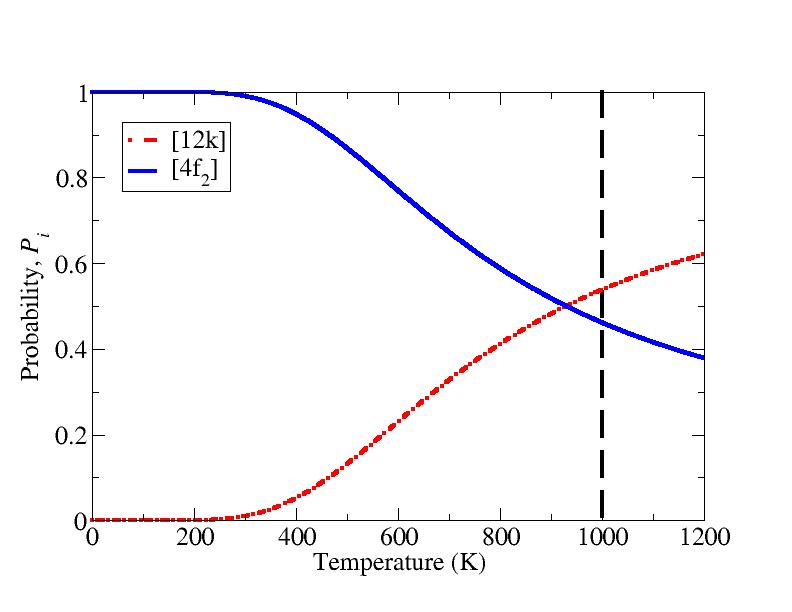}
  \end{tabular}
  \caption{\label{fig:Sn} Temperature dependence of the formation
    probability of different configurations of
    SrFe$_{11.5}$Sn$_{0.5}$O$_{19}$.  The configurations with
    negligible probability are not shown.  The vertical dotted line
    indicates the annealing temperature of 1000~K.  }
\end{figure}

\begin{figure}[tbp]
  \centering
  \begin{tabular}{c}
    \includegraphics[scale=0.35,keepaspectratio=true]{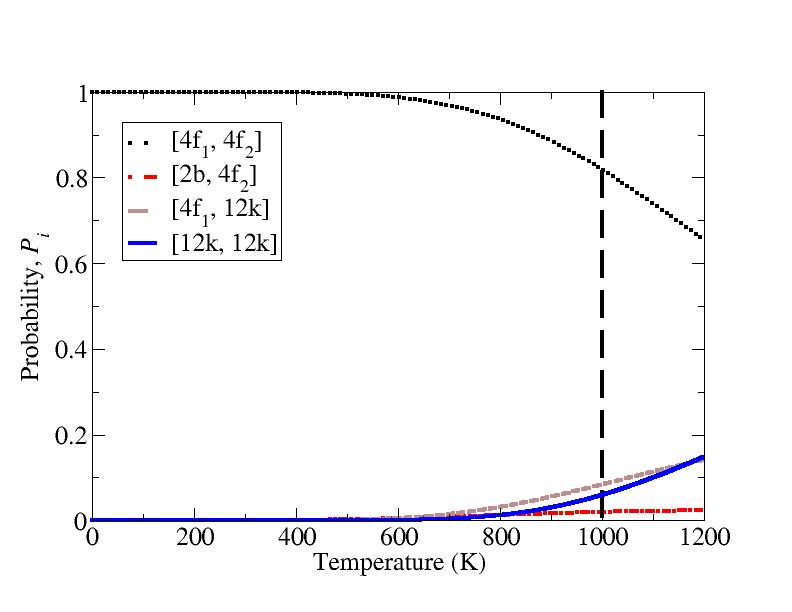}
  \end{tabular}
  \caption{\label{fig:ZnSn} Temperature dependence of the formation
    probability of different configurations of
    SrFe$_{11.0}$Zn$_{0.5}$Sn$_{0.5}$O$_{19}$.  The configurations
    with negligible probability are not shown. The vertical dotted line
    indicates the annealing temperature of 1000~K. }
\end{figure}

Fig.~\ref{fig:Zn} shows the variation of site occupation probability
with temperature for the inequivalent configurations of
SrFe$_{12-x}$Zn$_{x}$O$_{19}$ with $x=0.5$. In low temperature range
the $4f_1$ site is most likely to be occupied. However, as temperature
rises this probability falls while the site occupation probability of
$12k$ rises. This is due to small substitution energy difference
between $4f_1$ and $12k$ sites and higher multiplicity of the $12k$
site.  Similar behavior of site occupancy was seen in the case of
single Sn atom substitution as shown in Fig.~\ref{fig:Sn}. In the
elevated temperature regime, the site occupancy of the $4f_2$ falls,
while that of the $12k$ site rises.  In fact, at the annealing
temperature of SFO (1000~K), the occupation probability of the $12k$
site (54\%) becomes higher than that of the $4f_2$ site (46\%), which
is the most energetically stable substitution site at 0~K.  For the
Zn-Sn substitution (SrFe$_{11}$Zn$_{0.5}$Sn$_{0.5}$O$_{19}$), the
formation probability of $[4f_1,4f_2]$ is high in the low as well as
high temperature range. Other important configurations with
significant formation probability at high temperature (1000 K) are
$[4f_1, 12k]$ and $[12k, 12k]$ (Fig.~\ref{fig:ZnSn}).  Substituted Zn
atoms are most likely to occupy $4f_1$ and $12k$ sites, while Sn atoms
occupy $4f_2$ and $12k$ sites.

\begin{table*}[tbp]
  \caption{\label{tab:final} Weighted averages of physical properties
    of pure and substituted (Zn, Sn, Zn-Sn) strontium hexaferrite:
    volume of the unit cell ($V$), total magnetic moment
    ($m_{\text{tot}}$), saturation magnetization ($\sigma_s$),
    magnetocrystalline anisotropy energy ($E_a$), uniaxial magnetic
    anisotropy constant ($K_1$), and anisotropy field ($H_a$).}
  \centering
  \begin{ruledtabular}
    \begin{tabular*}{\hsize}{cccccccccccr}
      material & $\exptval{V}$ ($\AA^{3}$)
      & $\exptval{m_{\text{tot}}}$ ($\mu_{\text{B}}$) 
      & $\exptval{\sigma_{s}}$ (emu/g)
      &  $E_a$ (meV) & $K_1$ (kJ/m$^{3}$) & $H_a$ (kOe) \\
    \hline
    \SFO\     & 706.83 & 39.0 & 110.19 & 0.85 & 193.00 & 7.35 \\
    \SFZnO\   & 706.83 & 41.7 & 109.22 & 0.83 & 187.39 & 6.75 \\
    \SFSnO\   & 718.61 & 39.7 & 101.34 & 0.75 & 168.05 & 6.52 \\
    \SFZnSnO\ & 717.08 & 47.2 & 120.01 & 0.69 & 153.87 & 4.99 \\ 
    \end{tabular*}
  \end{ruledtabular}
\end{table*}

\begin{table*}[tbp]
  \caption{\label{tab:result} Comparison of calculated and
    experimental magnetic properties of \SFZSOx\ with $x=0$ and
    $x=0.5$: saturation magnetization ($\sigma_s$), uniaxial magnetic
    anisotropy constant ($K_1$), and anisotropy field ($H_a$).  The
    calculated values are for 0~K while the experimental values are
    measured at the room temperature.  Relative difference w.r.t.
    pure SFO ($x=0$) values are given in parentheses.  }
  \centering
  \begin{ruledtabular}
    \begin{tabular*}{\hsize}{ccccccc}
      & \multicolumn{2}{c}{$\sigma_s$ (emu/g)} 
      & \multicolumn{2}{c}{$K_1$ (kJ/m$^{3}$)} 
      & \multicolumn{2}{c}{$H_a$ (kOe)} \\
      \cline{2-3}
      \cline{4-5}
      \cline{6-7}
      & $x=0$ & $x=0.5$
      & $x=0$ & $x=0.5$
      & $x=0$ & $x=0.5$ \\
      \hline
      Exp.~\cite{FANG1998129, Ghasemi:2010} & 69.9 & 75.1 ($+7.4\%$) 
      & 280.9 & 210.7 ($-25.0\%$) & 18.7 &  12.5 ($-33.0\%$) \\ 
      Calc.~[This work] & 110.2 & 120.0 ($+8.2\%$)
      & 193.0 & 153.9 ($-20.0\%$) & 7.4 &  5.0 ($-32.0\%$) \\
      Calc.~\cite{Liyanage2013} & 113.1 & 127.2 ($+12.5\%$)
      & 190.0 & 100.0 ($-47.4\%$) & 7.1 &  3.0 ($-57.0\%$) \\
    \end{tabular*}
  \end{ruledtabular}
\end{table*}

The total magnetic moment $m_{\text{tot}}$ upon the substitution of a
single Zn and Sn atom at various Fe site of SFO is given in
Table~\ref{tab:SFZnOx0.5} and Table~\ref{tab:SFSnOx0.5}. It is evident
that the substitution at the minority spin sites ($4f_1$ and $4f_2$)
enhances the net magnetic moment, while moment is reduced for the
substitution at the majority spin sites ($12k$, $2a$, and $2b$).
Similar changes in the total magnetic moment were also noticed for
(SrFe$_{11}$Zn$_0.5$Sn$_{0.5}$O$_{19}$ substitution
(Table~\ref{tab:SFZnSnOx0.5}). The saturation magnetization $\sigma_s$
values are also listed.
 
$E_a$, $K_1$, and $H_a$ of various configurations are presented in
Table~\ref{tab:SFZnOx0.5}, \ref{tab:SFSnOx0.5} and
\ref{tab:SFZnSnOx0.5}. In these cases, substitutions at $2a$ site
seems to enhance the anisotropy values.  Table~\ref{tab:SFZnSnOx0.5}
also that the reduction of magnetic coercivity in \SFZnSnO\ as
experimentally observed \cite{FANG1998129} is mainly due to the
occupation of (Zn,Sn) pair at the ($4f_1$, $4f_2$) sites.  Although
the $[2b,4f_{2}]$ configuration, the one with second lowest
substitution energy at 0~K, has much lower $K_1$ and $H_a$ values,
its multiplicity is much lower compared to other low-energy
configurations and its formation probability grows merely to 2\%
(compared to 82\% for the $[4f_{1},4f_{2}]$ configuration) even at
1000~K.  This is a different behavior from the related systems
BaFe$_{12-x}$(Zr$_{0.5}$Zn$_{0.5}$)$_x$O$_{19}$ and
LaZnFe$_{11}$O$_{19}$, where the $2b$ site plays the main role in the
reduction of anisotropy \cite{Li:2000, Obradors:1985}.

Table~\ref{tab:final} shows the weighted average of physical
properties of pure and substituted (Zn, Sn, Zn-Sn) strontium
hexaferrite based on the formation probabilities at the annealing
temperature.  In all three cases of substituted SFO, the
$m_{\text{tot}}$ was estimated to be greater than that of pure SFO.
Biggest increase was found in the case of Zn-Sn pair substituted SFO.
A reduction in the values of $\sigma_{s}$ and anisotropy of
SrFe$_{11.5}$Zn$_{0.5}$O$_{19}$ and SrFe$_{11.5}$Sn$_{0.5}$O$_{19}$ is
in agreement with previous experimental studies on substituted barium
hexaferrite \cite{VINNIK2015, GONZALEZANGELES2004}.  The saturation
magnetization ($\sigma_s$) of SrFe$_{11.5}$Zn$_{0.5}$O$_{19}$ and
SrFe$_{11}$Zn$_{0.5}$Sn$_{0.5}$O$_{19}$ were higher than that of pure
SFO.  In Table~\ref{tab:result} we compare the magnetic properties of
Zn-Sn pair substituted SFO in this work with previously reported
computed and experimental data.  Results from the present work show an
increase of 8.2\% in the value of $\sigma_s$, while Ghasemi \etal
\cite{Ghasemi:2010} experimentally observed a similar increase of
7.4\%.  Quantities related to magnetic anisotropy viz. $K_1$, and
$H_a$ showed reduction compared to pure SFO.
SrFe$_{11}$Zn$_{0.5}$Sn$_{0.5}$O$_{19}$ showed a reduction of 20\% and
32\% in $K_1$ and $H_a$, respectively, while experimentally Fang \etal
observed a reduction of 25\% and 33\% in $K_1$ and $H_a$ values
\cite{FANG1998129}.  Table~\ref{tab:result} indicates that our present
results are in better agreement with experimental results compared to
the previous computed results that was obtained by considering the
ground state configurations at 0~K only. Thus, our model based on the
formation probability at the annealing temperature successfully
explains the reduction in anisotropy of Zn-Sn substituted SFO as well
as the increase in magnetization values.

\section{Conclusion}

Using first-principles total energy calculations based on density
functional theory, we calculated substitution energies of Zn, Sn, and
Zn-Sn pair substituted strontium hexaferrite.  These energy values
were used to determine site preferences of the substituted atoms at
0~K as well as at a high annealing temperature. The site occupation
probabilities or the formation probabilities of different
configurations were then used to estimate magnetic properties of
substituted SFO. We found that in SrFe$_{11.5}$Zn$_{0.5}$O$_{19}$, Zn
atoms prefer to occupy $4f_1$, $12k$, and $2a$ sites with occupation
probability of 78\%, 19\% and 3\%, respectively, while in
SrFe$_{11.5}$SnO$_{19}$, Sn atoms occupy the $12k$ and $4f_2$ sites
with occupation probability of 54\% and 46\%, respectively.  We
further showed that in SrFe$_{11}$Zn$_{0.5}$Sn$_{0.5}$O$_{19}$, the
pair of (Zn,Sn) atoms prefers to occupy the ($4f_1$, $4f_2$), ($4f_1$,
$12k$) and ($12k$, $12k$) sites with occupation probability of 82\%,
8\% and 6\%, respectively.  The results from our model based on the
formation probability were found to be in good agreement with recent
experimental observations showing enhancement of magnetization and the
reduction in anisotropy for Zn-Sn substituted strontium hexaferrites.

\section{Acknowledgment}

This research was supported by the Creative Materials Discovery
Program through the National Research Foundation of Korea (NRF) funded
by the Ministry of Science, ICT, and Future Planning
(2016M3D1A1919181).  It was also partly supported by the Center for
Computational Sciences (CCS) at Mississippi State University.  YKH's
work was partially supported by National Science Foundation (NSF)
under Award Number CMMI 1463301.  Computer time allocation has been
provided by the High Performance Computing Collaboratory (HPC$^2$) at
Mississippi State University.

\bibliography{SFO}

\end{document}